\documentclass[preprint,12pt]{elsarticle}

\usepackage{amssymb}

\journal{Physics Letter B}

\begin{document}

\begin{frontmatter}



\title{Correlations between the fragmentation modes and light charged particles emission}


\author[1]{Yingxun Zhang\corref{cor1}}
\ead{zhyx@ciae.ac.cn}
\cortext[cor1]{Corresponding author}
\author[1,2] {Chengshuang Zhou}
\author[1,2] {Jixian Chen}
\author[2] {Ning Wang}
\author[1] {Kai Zhao}
\author[1] {Zhuxia Li}

\address[1] {China Institute of Atomic Energy, P.O. Box 275 (10), Beijing 102413, P.R. China}
\address[2] {College of Physics and Technology,Guangxi Normal University, Guilin 541004,P.R. China}

\begin{abstract}
The correlations between the shape of rapidity distribution of the yield of light charged particles and the fragmentation
modes in semi-peripheral collisions for $^{70}$Zn+$^{70}$Zn, $^{64}$Zn+$^{64}$Zn and $^{64}$Ni+$^{64}$Ni at the beam energy of 35MeV/nucleon are
investigated based on ImQMD05 code. Our studies show there is interplay between the binary, ternary and multi-fragmentation break-up modes.
The binary and ternary break-up modes more prefer
to emit light charged particles at middle rapidity and give larger values of $R_{yield}^{mid}$ compared with
the multi-fragmentation break-up mode does. The reduced rapidity distribution for
the normalized yields of p, d, t, $^3$He, $^4$He and $^6$He and the corresponding values
of $R_{yield}^{mid}$ can be used to estimate the probability of multi-fragmentation break-up modes.
By comparing to experimental data,
our results illustrate that $\ge$40\% of the collisions events belong to the multi-fragmentation break-up mode for the reaction we studied.

\end{abstract}


\end{frontmatter}


Fragmentation mechanism, i.e. does a nucleus break up into several intermediate mass fragments (IMFs)
in intermediate energy heavy ion collisions (HICs) in a statistical (thermal) way or dynamical way,
has been a long-standing problem in nuclear physics, and it has attracted a lot of theoretical
and experimental efforts. Up to now, both the statistical and dynamical approaches can equally
well describe many observables in intermediate energy HICs, such as charge distributions,
mean kinetic energies of the emitted fragments and so on, but each of them fail to describe
some observables predicted by others, such as the isotope yields, isoscaling,
isospin diffusion, bimodality
\cite{Bondorf95,Gross90,Tsang, Souza96,Agosti96,William97,Ogul11,Aiche91,Hart98, Zhang06, Zhang07, Fever08,Amori09,Cardel12}.
It turns out that this problem is still open and the situation is very complicated.
One may not simply attribute the fragmentation mechanism to dynamical or statistical way,
and both approaches may be complement each other for low-intermediate energy heavy ion collisions.

Previous study in reference\cite{ZXLi91}
has shown there is interplay between the binary/ternary break-up (fusion-fission) modes
and multi-fragmentation mode, especially for the low-intermediate energy HICs (20-40MeV/nucleon).
The binary and ternary break-up \cite{Amori09,Cardel12, ZXLi91, Tian09, Filip05, SHudan12, Mcton10}
modes are mainly related to a slow process\cite{RSWang14,QHWu15} which can also be described in the statistical way,
and multi-fragmentation break-up mode is related to a fast process which can be described in dynamical way.
Special attentions have been paid to semi-peripheral collisions at low-intermediate energy for further understanding the fragmentation mechanism,
because it is the best way to study the interplay between fusion-fission and
multi-fragmentation \cite{Filip05,SHudan12,Mcton10,Souli03,Souli04,Zhang05,Kohley,Kohley12} process. In experiment, Kohley et al studied
the semi-peripheral collisions of
$^{70}$Zn+$^{70}$Zn, $^{64}$Zn+$^{64}$Zn, and $^{64}$Ni+$^{64}$Ni at the beam energy of 35MeV/nucleon\cite{Kohley}.
The data of rapidity distributions for normalized yields of isotopically light charged particles (LCPs), i.e. $Y(y_r)$/$Y(y_r=0)$,
of p, d, t, $^3$He, $^4$He and $^6$He, are reported.
Those data show a clear preference for emission around the middle rapidity region ($|y_r|<0.5$, where $y_r=\frac{y^{c.m.}}{y^{c.m.}_{proj}}$),
especially for more neutron-rich LCPs due to the isospin drift and diffusion\cite{Baran05, Zhang05}.
However, the experimental results on the reduced rapidity distributions of the normalized yields of proton and $^3$He
tend to have an increased forward rapidity yield relative to the Stochastic Mean Field model(SMF)\cite{Maria04,Rizzo}
calculations with both soft and stiff symmetry energy cases. Those discrepancies could be attributed to the decay of quasi-projectile (QP) at
later stages of the reaction \cite{Kohley}. Recent studies for $^{124}$Sn+$^{64}$Ni show that the parallel velocity distribution for the yields of Z=3-6 fragments can be described after coupling the sequential decay to the transport models\cite{Filip12}, but there still have little differences between data and theory. It seems to us that the statistical decay is not enough to explain the discrepancy on the rapidity distribution of LCPs between theory and experiment, and we still need to understand it in the dynamical approaches besides sequential decay. By the way, the symmetry energy play important roles in determination of the isotope distribution of emitted fragments in transport model because its values directly related to the thickness of neutron skin and the borderline of nuclear chart\cite{XHFan15,QHMo15, XYQu13} in theoretical prediction.

In this paper, we investigate the dynamical correlations between the fragmentation modes and LCPs emission by
analyzing the rapidity distributions of the yields for LCPs and the branching ratio of different break-up mode for semi-peripheral collisions of
$^{70}$Zn+$^{70}$Zn, $^{64}$Zn+$^{64}$Zn, and $^{64}$Ni+$^{64}$Ni at the beam energy of 35MeV/nucleon
by using ImQMD05 code. The influence of different impact parameters and symmetry energy are also discussed. Since the main goal of this work is to investigate the correlation
between the fragmentation modes and LCPs emission, there is no attempt to obtain the best fit to the data.



Within the ImQMD05, nucleons are represented by Gaussian wavepackets and the nucleonic mean fields acting these wavepackets are derived from an energy density functional. The potential energy U that includes the full Skyrme potential energy with just the spin-orbit term omitted and the Coulomb energy reads as:
\begin{equation}
U=U_{\rho,md}+U_{coul}.
\end{equation}
The nuclear contributions are represented in a local form with
\begin{equation}
 U_{\rho,md}=\int u_{\rho,md}d^3r
\end{equation}
and
\begin{eqnarray}
u_{\rho}=&&\frac{\alpha}{2}\frac{\rho^{2}}{\rho_{0}}+\frac{\beta}{\eta+1}\frac{\rho^{\eta+1}}{\rho^{\eta}_{0}}+\frac{g_{sur}}{2\rho_{0}}(\nabla \rho)^2 \nonumber\\
&&+\frac{g_{sur,iso}}{\rho_{0}}(\nabla (\rho_{n}-\rho_{p}))^2 \nonumber\\
&&+\frac{C_{s}}{2}(\frac{\rho}{\rho_{0}})^{\gamma_i}\delta^{2}\rho+g_{\rho\tau}\frac{\rho^{8/3}}{\rho_{0}^{5/3}} \label{urho}
\end{eqnarray}
Here, $\delta$ is the isospin asymmetry. $\delta=(\rho_n-\rho_p)/(\rho_n+\rho_p)$, $\rho_n$ and $\rho_p$ are the neutron and proton densities, respectively. A symmetry potential energy density of the form $\frac{C_s}{2}(\frac{\rho}{\rho_0})^{\gamma_i}\delta^2\rho$ is used in the following calculations. The energy density associated with the mean-field momentum dependence is represented by
\begin{eqnarray}
u_{md}=\frac{1}{2\rho_{0}}\sum_{N_1,N_2} \frac{1}{16\pi^{6}}\int d^{3}p_{1}d^{3}p_{2}f_{N_{1}}(\vec{p}_1)f_{N_{2}}(\vec{p}_2)\nonumber\\
1.57[\ln(1+5\times 10^{-4}(\Delta p)^2)]^2\;. \label{eq:umd}
\end{eqnarray}
where $f_N$ are nucleon Winger functions, $\Delta p=|\vec{p_1}-\vec{p_2}|$, the energy is in MeV and momenta are in MeV/c.
The resulting interaction between wavepackets is described in Ref. \cite{Aiche87}.
In this work, the $\alpha=-356$MeV, $\beta=303$MeV, $\eta=7/6$, and $g_{sur}=19.47$ $MeVfm^2$, $g_{sur,iso}=-11.35$ $MeVfm^2$, $C_s=35.19$MeV,
and $g_{\rho\tau}=0$MeV. These calculations use isospin-dependent in-medium nucleon-nucleon scattering cross sections
in the collision term and the treatment of Pauli blocking effect is described in \cite{Zhang05, Zhang06, Zhang07}.
Fragments are formed in QMD approaches due to the
A-body correlation dynamics and these correlations can be mapped onto the asymptotic final fragments by using the cluster recognition methods,
such as minimum spanning tree algorithm (MST) and so on\cite{Dorso93,Puri00,Goyal11}.
In this work, the clusters are recognized by means of the isospin dependent cluster recognition method
(iso-MST)\cite{Zhang12c} which is an extended version of MST method and refines the description on the
yield of neutron-rich LCPs, especially for the neutron-rich reaction systems.

Figure 1 presents the time evolution of the density contour plots for typical event of reaction $^{64}$Zn+$^{64}$Zn at E$_{beam}$=35MeV/nucleon and b=4fm.
The results are obtained with $\gamma_i=0.5$. As shown in figure 1, the projectile and target touch around 60fm/c and
form a low density neck region through which the nucleons transfer between the projectile and target. The compressed region reaches highest
compress density at 110fm/c. The projectile and target separate after 160fm/c, whereupon the low-density neck
connecting them ruptures into fragments. Two excited residues then continue along their paths and emit lighter fragments.
The fluctuations which automatically appears in the QMD type models lead the system breaks via different break-up modes ,
such as binary, ternary and multi-fragmentation break-up modes as shown in the Figure 1 (f1), (f2) and (f3), with different probabilities.
Since the different break-up modes obviously lead to
different emission patterns as well as the different angular and rapidity distributions of light charged particles
(LCPs) and fragments (for example, Figure 1 (f1-f3)), one can expect that the branching ratios of binary, ternary and multi-fragmenation break-up modes directly
influence final results on the shape of
rapidity distribution of the yields of LCPs.

\begin{figure}[htbp]
\centering
\includegraphics[angle=270,scale=0.40]{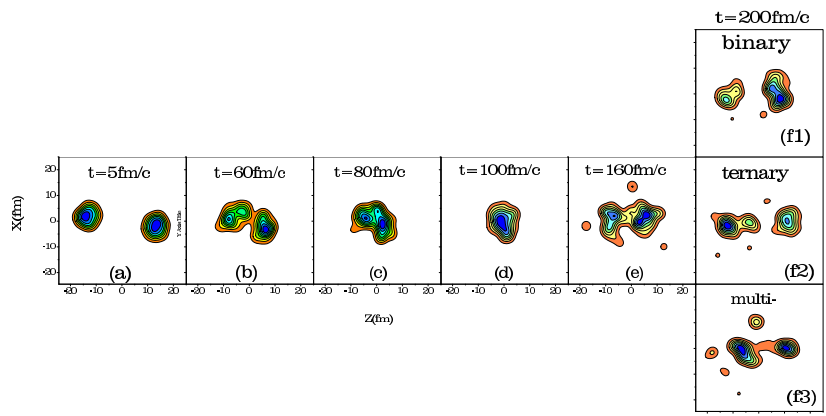}
\setlength{\abovecaptionskip}{50pt}
\caption{(Color online)Time evolution of the density contour plots for $\mathrm{^{64}Ni+ ^{64}Ni}$ at $E_{beam}$=35 MeV/nucleon for b=4fm from typical events which are calculated with $\gamma_i=0.5$.}\label{ref-Fig.1}
\setlength{\belowcaptionskip}{10pt}
\end{figure}

In Figure 2(a), we present the branching probability of different break-up modes for $^{70}$Zn+$^{70}$Zn at
E$_{beam}$=35MeV/nucleon and b=4fm with
$\gamma_i$ =2.0 (solid symbols) and 0.5 (open symbols). The multi-fragmentation break-up mode is defined by M(Z$\ge$3)$\ge$4
in this work, where M(Z$\ge$3) is the multiplicity
for fragments with Z$\ge$3. The binary and ternary break-up modes mean the M(Z$\ge$3)=2 and M(Z$\ge$3)=3, respectively. Our calculations
with $\gamma_i$=2.0 ($\gamma_i$=0.5) show that $\sim$18\% ($\sim$19\%) events for semi-peripheral $^{70}$Zn+$^{70}$Zn
collisions belong to binary mode, $\sim$36\% ($\sim$32\%) events for ternary mode. The rest, almost $\sim$46\% ($\sim$49\%)
of total events, belong to the multi-fragmentation break-up mode as shown in Figure 2(b). Similar behaviors can be observed
for $^{64}$Zn and $^{64}$Ni system. The values of the probability for M(Z$\ge$3)$\ge$4 slightly depend on the mass of
reaction system, those for $^{70}$Zn are the largest for both symmetry energy cases.
\begin{figure}[htbp]
\centering
\includegraphics[angle=270,scale=0.35]{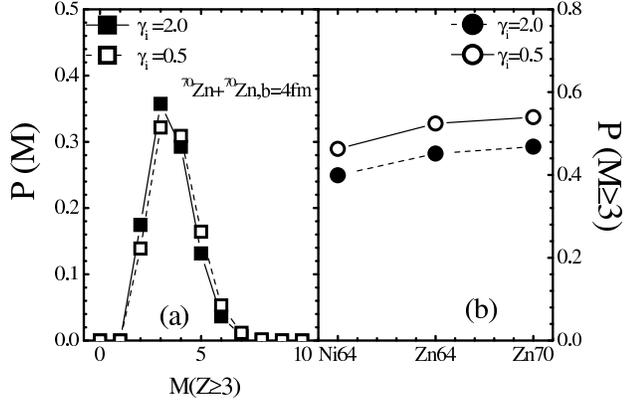}
\setlength{\abovecaptionskip}{35pt}
\caption{\label{ref-Fig.2}(Color online) (a) The calculated results on possibility distribution of multiplicity for fragments with $Z\geq 3$ ($M(Z\geq 3)$) for $\mathrm{^{70}Zn+^{70}Zn}$ at E$_{beam}$=35 MeV/nucleon and b=4fm, with $\gamma_i=2.0$ (solid symbols) and 0.5 (open symbols); (b) The possibilities of multi-fragmentation modes for $^{70}Zn+^{70}Zn$, $^{64}Zn+^{64}Zn$ and $^{64}Ni+^{64}Ni$ are for $\gamma_i=2.0$ (solid symbols)and 0.5 (open symbols).}
\setlength{\belowcaptionskip}{0pt}
\end{figure}

To understand the correlations between the break-up modes and LCPs emissions,
we plot the rapidity distributions for the yields of LCPs corresponding to
three kinds of break-up modes, binary (square symbols), ternary (circle symbols) and multi-fragmentation
(triangle symbols) in Figure 3. Fig. 3 (a) and (b) are for neutron-poor particle, Y($^3$He), (c) and (d) are for
neutron-rich particles, Y($^6$He). Left panels are for $\gamma_i$=2.0 and right panels are for $\gamma_i$=0.5.
The rapidity distributions for the yields of $^{3, 6}$He are normalized to per event.
It is clear that the binary and ternary modes tend to produce more $^{3}$He and $^{6}$He at midrapidity
than that multi-fragmentation break-up mode does. The difference of Y($^{6}$He) between the binary, ternary modes
and multi-fragmentation break-up mode is larger due to stronger isospin migration compared with that for Y($^{3}$He).
For example, for $\gamma_i$=2.0, the yield of $^3$He at $y_r\sim0$ from binary mode is 35\% larger than that
from multi-fragmentation mode, but the yield of $^6$He at $y_r\sim0$ from binary mode is 70\% larger than that from
multi-fragmentation break-up mode. Consequently, one can expect that the different break up modes lead to
different shape of rapidity distribution of LCPs.
The values of $R_{yield}^{mid}$, which is defined by the yield ratio between the mid-rapidity and projectile rapidity,
\begin{equation}
R^{mid}_{yield}=\frac{2\cdot Y(0.0\leq y_r \leq0.5)}{Y(0.5\leq y_r \leq1.5)}
\end{equation}
can quantitatively reflect the shape of rapidity distribution of the yields of LCPs, i.e. $Y(y_r)/Y(y_r=0)$,
as well as the degree of the preference of mid-rapidity emission for LCPs \cite{Kohley}.
Based on above definition, one expects $R^{mid}_{yield}>1$ if there is a preference for emission around the middle rapidity.
The narrower the shape of rapidity distribution of the normalized yield of LCPs is, the larger the $R^{mid}_{yield}$ is.
As results, the $R_{yield}^{mid}$ values obtained with binary break-up mode are larger than that
with multifragmentation break-up mode, especially
for the neutron-rich LCPs in dynamical approaches. For example, the value of $R_{yield}^{mid} (^6He)$ is 2.9
in binary break-up mode events, 2.5 in ternary break-up mode events and 2.4
in multi-fragmentation break-up mode events calculated with $\gamma_i$=2.0. This relationship
between the $R_{yield}^{mid}$ values and break-up modes is also valid
for $\gamma_i$=0.5, but their absolute values are different. The $R_{yield}^{mid} (^6He)$ obtained
with $\gamma_i$=0.5 is 2.1 in binary break-up mode events, 1.8 in ternary break-up mode events and 1.8 in multi-fragmentation
break-up mode events.

\begin{figure}[htbp]
\centering
\includegraphics[angle=270,scale=0.35]{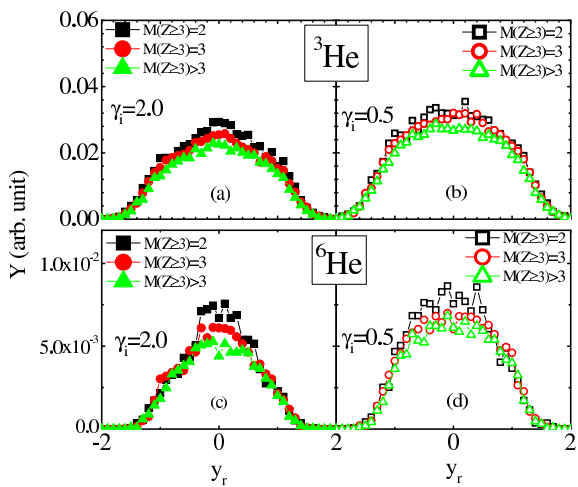}
\setlength{\abovecaptionskip}{35pt}
\caption{\label{ref-Fig.3}(Color online) (a) and (b) are the reduced rapidity ($y_r$) distribution for the yield of $^3He$, Y($^3He$), with binary (square symbols), ternary (circle symbols) and multi-fragmentation (triangle symbols) break-up modes. (c) and (d) are for Y($^6He$). (a) and (c) are the results with $\gamma_i=2.0$, (b) and (d) are for $\gamma_i=0.5$. All of those results are for $\mathrm{^{70}Zn+^{70}Zn}$ at E$_{beam}$=35 MeV/nucleon for b=4fm.}
\setlength{\belowcaptionskip}{0pt}
\end{figure}

Now, let's compare the calculated results of $Y(y_r)/Y(y_r=0)$ to the experimental data\cite{Kohley}.
100,000 events are performed for each reaction at
different impact parameters. As an example, we present the $Y(y_r)/Y(y_r=0)$ for p, d, t, $^3$He, $^4$He and $^6$He,
for neutron-rich system $^{70}$Zn+$^{70}$Zn at b=1, 6 and 8 fm with $\gamma_i$=2.0 in Figure 4.
Our calculations show that there is no strong
dependence on the impact parameters in the range of experimental centrality $b/b_{max}\cong 0.33-0.66$\cite{Kohley}, i.e. b$\cong$3-6fm. Since the published data of $Y(y_r)/Y(y_r=0)$ \cite{Kohley} is for $^{64}$Ni+$^{64}$Ni,
we only plot the $Y(y_r)/Y(y_r=0)$ for p, d, t, $^3$He, $^4$He and $^6$He,
for $^{64}$Ni+$^{64}$Ni at b=4fm with $\gamma_i$=0.5 (open circles) and $\gamma_i$=2.0 (solid circles) in Figure 5 to get the physics insights. The data is plotted as stars\cite{Kohley}.
Both our calculations and data show that the width of distribution
decreases with the mass of LCP increasing, because the Fermi motion of nucleons has larger impacts on
the motion of lighter LCPs than that on heavier LCPs. Both theoretical and experimental results also
clearly show that the width of $^3$He is smaller than that of $^3$H due to the isospin migration effects.
Since the reaction systems calculated with $\gamma_i$=0.5 produce larger probabilities of multi-fragmentation mode,
and break up into more LCPs over the whole rapidity region than that with $\gamma_i$=2.0,
the $Y(y_r)/Y(y_r=0)$ for LCPs calculated with $\gamma_i$=0.5 are wider than that with $\gamma_i$=2.0,
especially for neutron rich LCPs. By comparing the simulated results to the data,
one can find that the ImQMD05 calculations with stiffer symmetry energy case can well reproduce
the data at forward rapidity region ($y_r>0$) for all p, d, t, $^3$He, $^4$He and $^6$He particles.
It suggests that branching probabilities for multi-fragmentation break-up modes should
be around $\sim$46\% for $^{64}$Ni+$^{64}$Ni as gained in this work.
For the backward rapidity region ($y_r<0$), the obvious difference between the calculated results and the data is caused by lacking the efficiency for detection of LCPs at the backward \cite{Kohley}
in calculation.


\begin{figure*}[htbp]
\centering
\includegraphics[angle=270,scale=0.4]{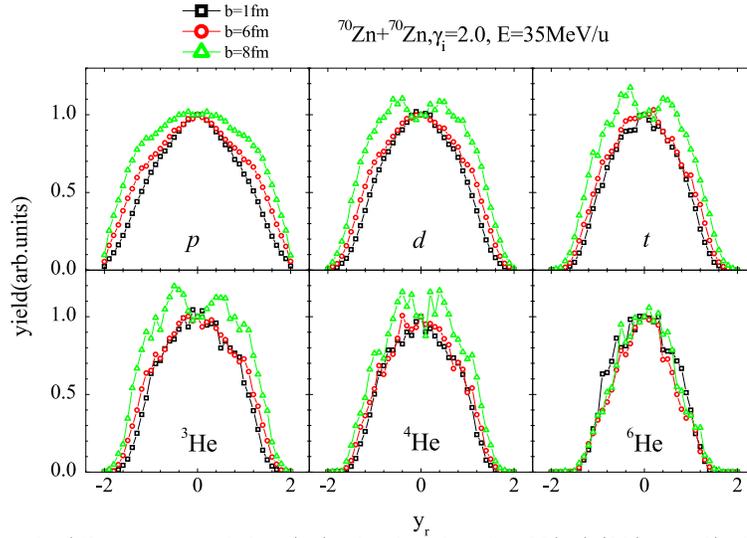}
\setlength{\abovecaptionskip}{35pt}
\caption{(Color online)Reduced rapidity ($y_r$) distribution for $Y(y_r)/Y(y_r=0)$,
for \emph{p}, \emph{d}, \emph{t}, $^{3}He$, $^{4}He$ and $^{6}He$ for $\mathrm{^{70}Zn+^{70}Zn}$ at impact parameter b=1, 6, 8fm. The calculations are obtained with $\gamma_i$=2.0}\label{ref-Fig.4}
\setlength{\belowcaptionskip}{0pt}
\end{figure*}

\begin{figure*}[htbp]
\centering
\includegraphics[angle=270,scale=0.4]{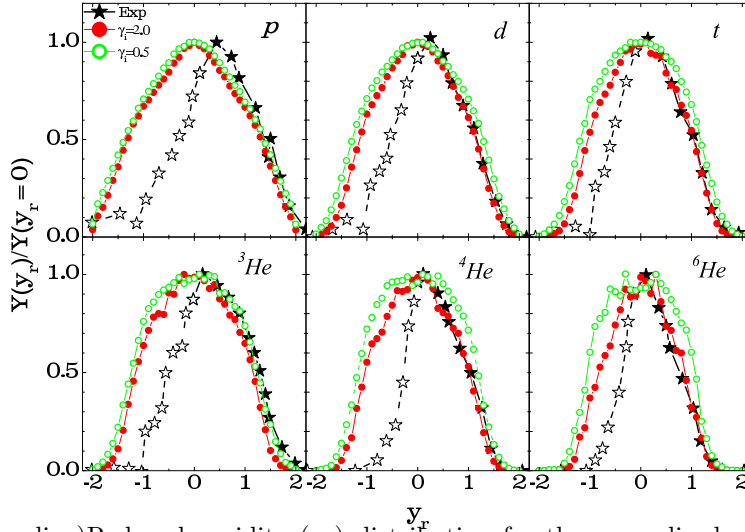}
\setlength{\abovecaptionskip}{35pt}
\caption{(Color online)Reduced rapidity ($y_r$) distribution for the normalized yields of \emph{p}, \emph{d}, \emph{t}, $^{3}He$, $^{4}He$ and $^{6}He$, i.e. $Y(y_r)/Y(y_r=0)$,
for $\mathrm{^{64}Ni+^{64}Ni}$ at 35MeV/nucleon and b=4fm.
The experimental data is shown as the stars. The ImQMD05 calculation for $\gamma_i=0.5$
is shown as the open circles and the solid circles are for $\gamma_i=2.0$.
Each distribution is normalized with its yield at $y_r=0$.}\label{ref-Fig.4}
\setlength{\belowcaptionskip}{0pt}
\end{figure*}

In order to quantitatively understand the influence of the density dependence of symmetry energy and sequential decay on
the shape of rapidity distribution of yields for LCPs, we investigate the $R_{yield}^{mid}$
values of the LCPs as a function of the product of charge and mass
(AZ=1 for $^1$H, AZ=2 for $^2$H, AZ=3 for $^3$H, AZ=6 for $^3$He, AZ=8 for $^4$He and AZ=12 for $^6$He)
for three reaction systems
$^{64}$Zn +$^{64}$Zn,$^{64}$Ni +$^{64}$Ni and $^{70}$Zn+$^{70}$Zn.
A series calculations with different density dependence of symmetry potential,
i.e. $\gamma_i$=0.5, 0.75, 1.0, 2.0 at b=4fm are performed.
As shown in Fig.6, the ImQMD05 calculations can reasonably reproduce the
data of $R_{yield}^{mid}$ as a function of AZ for $^{64}$Zn+$^{64}$Zn,
and qualitatively reproduce the trends of $R_{yield}^{mid}$ for neutron-rich reaction systems
$^{64}$Ni+$^{64}$Ni and $^{70}$Zn+$^{70}$Zn. The calculated results show the values of $R_{yield}^{mid}$
for neutron-rich isotopes, such as $^{6}$He, clearly depend on the density dependence of symmetry energy.
Since the calculations with stiffer symmetry energy have stronger isospin migration effects and predict
smaller probabilities of multi-fragmentation mode, it predicts larger $R_{yield}^{mid}$ values.
The calculations with symmetry energy ($\gamma_i\ge0.75$) reasonably reproduce the
data of $R_{yield}^{mid}$ as a function of AZ for $^{64}$Zn+$^{64}$Zn.
But for the neutron-rich reaction systems $^{64}$Ni+$^{64}$Ni and $^{70}$Zn +$^{70}$Zn, calculated results of $R_{yield}^{mid}$ for $^6$He are obviously smaller than the data whatever the form of symmetry energy is used in the model.
The possible reason is that the probabilities of binary and ternary
break-up modes in neutron-rich reaction systems
are underestimated in the QMD approaches than observed in experiments, especially for neutron rich systems. It may be caused by lacking the fine structure effects of neutron-rich nuclei (such as neutron skin, stability
of lighter neutron-rich nuclei) which may become more and more important with the beam energy decreasing in HICs simulations. Since the secondary decay also
plays important roles on the yields of LCPs, such as p, d, t, $^3$He,
$^4$He and $^6$He, and thus it may influences the ratio observables , such as $Y(y_r)/Y(y_r=0)$ and $R_{yield}^{mid}$. We have simulated
decays of fragments created in the collisions using the GEMINI
code \cite{Charity88} to examine the influence of sequential decays. Sequential decays enhance the yields of LCPs, but such effects are largely
suppressed in the ratio observables, $Y(y_r)/Y(y_r=0)$ and $R_{yield}^{mid}$ except for $^4$He. However, the exact effects after coupling the sequential decay to transport codes need to be further explored in the furture.

Moreover, the behaviors of $R_{yield}^{mid}$ as a function of AZ for He elements are different from the
SMF predictions (left triangles in Fig.6) in which a weakly AZ dependence of the $R_{yield}^{mid}$
for He isotopes was obtained \cite{Kohley}.
The possible reasons are: many-body correlations in the QMD models cause
large fluctuation, and it enhances the probability of multi-fragmentation break-up modes, and produces copious fragments
and light particles
over the whole rapidity region in the QMD simulations. Fast fragmentation dynamics inhibits nucleon
exchanging and charge equilibration due to the many-body correlations and strict Pauli blocking.
As a consequence, the neutron-rich He tend to locate at mid-rapidity than neutron-poor LCPs,
and thus, the larger the isospin asymmetry of He is, the larger the $R_{yield}^{mid}$ is.

\begin{figure}[tbp]
\centering
\includegraphics[angle=90,scale=0.50]{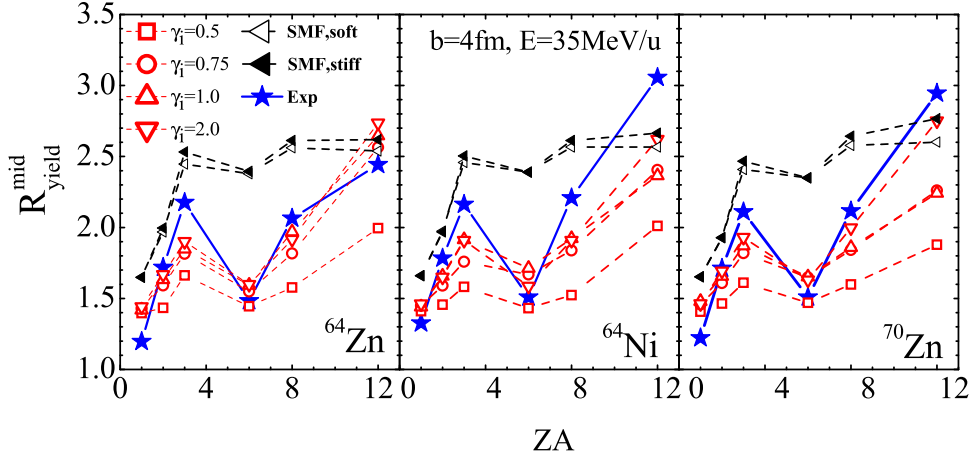}
\setlength{\abovecaptionskip}{10pt}
\caption{\label{ref-Fig.5}(Color online)$R_{yield}^{mid}$ values as
a function of the charge times mass (ZA) for \textit{p} (ZA=1), \textit{d} (ZA=2),  \textit{t} (ZA=3), $^3He$ (ZA=6),  $^4He$ (ZA=8), $^6He$ (ZA=12). The open symbols are the results obtained with ImQMD05 for $\gamma_i=0.5, 0.75, 1.0$ and $2.0$.
Left triangles are the results from SMF calculations and the solid stars are the data from \cite{Kohley}.}
\setlength{\belowcaptionskip}{0pt}
\end{figure}

In summary, we have studied the LCPs emissions in three reaction systems $^{70}$Zn+$^{70}$Zn, $^{64}$Zn+$^{64}$Zn and $^{64}$Ni+$^{64}$Ni
at the beam energy of 35MeV/nucleon and b=4fm to explore the fragmentation mechanism and relationship
between the break-up modes and rapidity distribution of
LCPs with ImQMD05. Our studies show there is interplay between the binary, ternary and multi-fragmentation break-up modes.
The binary and ternary break-up modes more prefer to produce LCPs at mid-rapidity, and it leads to a narrower
reduced rapidity distribution for the yield of LCPs than the multi-fragmentation break-up mode does. Based on the correlations between the fragmentation mode and LCPs emissions, one can estimate the probability of multi-fragmentation break-up modes by studying the reduced rapidity distribution for
the normalized yields of p, d, t, $^3$He, $^4$He and $^6$He and the corresponding values
of $R_{yield}^{mid}$. Current analysis suggests that reproducing the data of Y($y_r$)/Y($y_r=0$) for p, d, t, $^3$He, $^4$He and $^6$He and
corresponding $R^{mid}_{yield}$ in theory
requires that $\ge$40\% of the collisions events belong to the multi-fragmentation break-up mode in dynamical approaches. This values could also be slightly modified after including the tensor force and spin-orbit interaction, which were found to influence the dissipation dynamics at lower energy\cite{GFDai14}, in transport model.
Furthermore, experimental measurements/reanalysis on
the correlation between the probability of multi-fragmentation break-up mode and $R^{mid}_{yield}$ would give more
comprehensive information on the reaction mechanism, and it will be also helpful for tightly constraining the symmetry
energy with HICs in the future.


\textbf{Acknowledgements}
Yingxun Zhang thanks the Prof. M.Colonna on helpful communications.
This work has been supported by Chinese National Science Foundation under Grants
(11475262,11422548,11375062,11375094), the 973 program of China No. 2013CB834404.








\end{document}